\documentclass{qjmam}
\usepackage{amssymb}
\usepackage{amsmath}
\graphicspath{{./figures/}}
\usepackage{cleveref}
\usepackage{color}
\usepackage{calrsfs}
\usepackage[symbol]{footmisc}

\startpage{1}
\yr{2021}
\vol{52}
\issue{1}

\newcommand{\pfr}[2]{\ensuremath{\frac{\partial #1}{\partial #2}}}

\newcommand{\red}[1]{\textcolor{black}{#1}}

\newcommand{\ared}[1]{\textcolor{black}{#1}}  

\newcommand{\vect}[1]{\mathbf{#1}}

\newcommand{\beq}{\begin{equation}}
\newcommand{\eeq}{\end{equation}}

\DeclareMathAlphabet\mathbit
    \encodingdefault\rmdefault\bfdefault\itdefault
\DeclareOldFontCommand{\bi}{\normalfont\bfseries\itshape}{\mathbit}

\newcommand{\be}{\begin{equation}}
\newcommand{\ee}{\end{equation}}

\def\fakebold#1{\relax\ifvmode\leavevmode\fi%
\ifmmode%
\setbox0=\hbox{$#1$}%
\else%
\setbox0=\hbox{#1}%
\fi%
\kern-.02em\copy0 \kern-\wd0%
\kern .04em\copy0 \kern-\wd0%
\kern-.0125em\raise.02em\box0%
}%

\renewcommand{\geq}{\geqslant}
\renewcommand{\leq}{\leqslant}

\begin{document}

\title[{Vortices in radial stagnation flows}] {Steady axisymmetric vortices in radial stagnation flows}

\author[P.~Rajamanickam \& A.~D.~Weiss] {Prabakaran Rajamanickam}

\address{Department of Aerospace Engineering, Auburn University,\\
Auburn, AL {\rm 36849}, USA}

\extraauthor{Adam D. Weiss}

\extraaddress{ATA Engineering, Inc., San Diego, CA {\rm 92128}, USA}

\received{\recd . \revd }

\maketitle

\eqnobysec

\begin{abstract} 
A class of axisymmetric vortex solutions superposed upon radial stagnation flows is described. The new vortex solutions generalize the classical Burgers' vortex \ared{and Sullivan's vortex solutions} in the presence of a volumetric line source at the symmetry axis, \ared{the former} \red{approaching the Burgers' vortex sheet when the source strength becomes very large}. \ared{The stability of the generalized Burgers' vortex is studied}. In a different manner from \ared{the classical solution}, the \ared{generalized Burgers'} vortices are found to be unstable for two-dimensional disturbances when the vortex Reynolds number is increased above a critical value, for a fixed strength of the volumetric source. 
\end{abstract}

\section{Introduction}
While flows of the type describing stretched axes or, planes are extensively studied in the literature, studies on stretched cylindrical surfaces are quite limited perhaps because the latter flows are not as commonly observed as the former and are often realized only in laboratory experiments or special manufacturing processes. We shall refer to these latter flows as radial stagnation flows since stretched cylindrical surfaces are created by having a radially incoming flow towards the stagnation cylinder (defined in the same sense as one defines a stagnation surface in potential flow theory).

Stretched cylindrical surfaces are created by having long concentric, porous cylinders and injecting fluid, respectively, inwards and outwards from the inner and outer cylinder. Supposing that the radius of the inner and outer cylinder be denoted by $R_1$ and $R_2$ and the corresponding injection velocities, taken here to be constants, by $U_1$ and $U_2$, the volumetric flow rate per unit axial distance of the two cylinders can be written as $2\pi R_1U_1$ and $2\pi R_2 U_2$, respectively. Such an experimental setup, referred to as the opposed tubular burner, was first constructed by Hu \textit{et.~al.}~\cite{hu2007experimental} to investigate so-called tubular flames~\cite{ishizuka2013tubular} that offer a simple one dimensional setup to study the dependencies of strain and curvature on laminar flames. If the two cylinders are placed far away from each other and their volumetric flow rates are comparable, then the stagnation cylinder is expected to lie equally far away from both the cylinders. It turns out, in this case, a local description of the flow field can be described in the neighborhood of the cylindrical stagnation plane. Such descriptions must include only the gross features (like $R_1$, $U_1$, etc.,) of the cylinders, but not on the details adjacent to the cylindrical walls.

We choose cylindrical coordinates $(r^*,\theta,z^*)$ with corresponding velocity components $(v_r^*,v_\theta^*,v_z^*)$. Following the earlier work~\cite{seshadri1978laminar} for opposed jet burners, Wang \textit{et.~al.}~\cite{wang2006stretch} provided a large Reynolds number description for the opposed tubular burners where the function $r^*v_r^*$ was expressed in terms of the sine function. Expanding this solution in the neighbourhood of the stagnation surface, we immediately find the local description for the flow field, in the absence of any azimuthal motion, as
\begin{equation}
    v_r^* = -kr^* + \frac{q}{r^*}, \qquad v_z^* = 2 kz^* \label{vrvz}
\end{equation}
where $q\sim R_1U_1>0$ and $k\sim U_2/R_2>0$, in which $k$ can be interpreted as the strain rate imposed on the stagnation surface. This local description also emerges from the exact solution of the Navier-Stokes equations, discovered by Wang~\cite{wang1974axisymmetric} for a solid cylinder and Cunning \textit{et.~al.}~\cite{cunning1998radial} for a porous cylinder, placed in an axially straining flow. The location of the stagnation surface $r^*=\sqrt{q/k}$ occurs where the radial velocity vanishes. 

Often in the tubular combustion experiments, a swirling motion is introduced by allowing the radially incoming gas to pass through tangential slots~\cite{ishizuka2013tubular,ishizuka1993characteristics}. Interestingly enough, the vortex motion superposed on the velocity components~\eqref{vrvz} has not been previously considered, to the authors' knowledge, and serves as a new exact solution of the Navier-Stokes equations. It is the object of this paper to describe the structure and stability of this new vortex solution. It is also evident that when $q=0$, the vortex must become the Burgers' vortex. Before jumping into the vortex solution, it is of interest to point out another plausible application related to combustion. The solution can also be used to model the burning of a tall, wooden flagpole in a fire whirl~\cite{forman}, in which $q$ will be related to the gasification rates of the wooden fuel and the incoming flow will be driven ultimately by the fire whirl and the plume developing above.

\section{The axisymmetric vortex solution}
\label{sec:vortexsoln}

Consider an axisymmetric vortex superposed on the radial stagnation flow~\eqref{vrvz}, with a circulation strength $2\pi\tilde\Gamma$ found at large radial distances. Let all physical variables be non-dimensionalized from here onwards, using as time scales $k^{-1}$ and length scales $\sqrt{\nu/k}$, where $\nu$ is the kinematic viscosity of the fluid. The non-dimensionalization will introduce two parameters into the problem
\begin{equation}
    Q=\frac{q}{\nu} \quad \text{and} \quad R =\frac{\tilde\Gamma}{\nu}
\end{equation}
where $Q$ is the Reynolds number of the line source and $R$ is the vortex Reynolds number.

\begin{figure}
\centering
\includegraphics[width=0.6\textwidth]{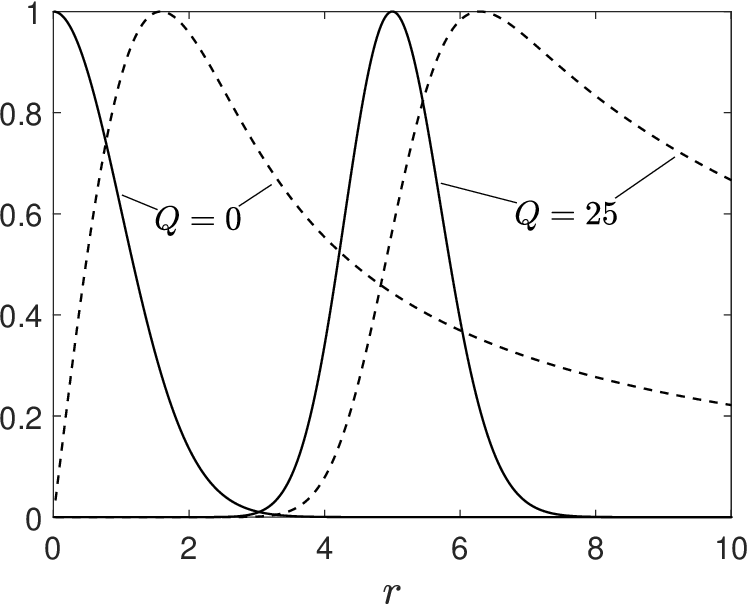}\\
\caption{The vorticity (solid curves) and azimuthal velocity (dashed curves) profiles for $Q=0$ and $Q=25$. \ared{Profiles have been normalized by their respective maximum value}.}
\label{fig:vortex}
\vspace{-0.15in}
\end{figure}

The vortex solution can be obtained from the Navier-Stokes equations by assuming
\begin{equation}
    v_r=-r + \frac{Q}{r}, \qquad v_\theta= \frac{R}{r}g(r), \qquad v_z=2z \label{vrvtvz}
\end{equation}
 where the equation satisfied by $g(r)$ is given by the azimuthal momentum equation
\begin{equation}
   \frac{d^2g}{dr^2} + \left(r-\frac{1+Q}{r}\right)\frac{dg}{dr}=0.
\end{equation}
The solution that satisfies the boundary conditions $g(0)=0$ and $g(\infty)=1$ is
\begin{equation}
    g(r) = 1-\frac{\Gamma(1+Q/2,r^2/2)}{\Gamma(1+Q/2)} \label{gsol}
\end{equation}
in which $\Gamma$ represents the gamma function. The non-zero vorticity component lying along the $z$ direction is
\begin{equation}
    \omega_z = \frac{R r^Q e^{-r^2/2}}{2^{Q/2}\Gamma(1+Q/2)} . \label{wz}
\end{equation}
This vorticity field is the required generalization of the Burgers' vortex in the presence of a volumetric line source at the axis. The vorticity reaches its peak value at the stagnation surface $r=\sqrt{Q}$, as shown in Figure~\ref{fig:vortex} for $Q=0$ and $Q=25$. As can be seen, the vortex core lies in an annular region centered at the stagnation surface for $Q\neq 0$.

The corresponding vortex solution when the line source is replaced by a porous cylinder of radius $R_1$ with injection velocity $U_1$ is given in Appendix A.

\subsection{Limiting form for large $\sqrt Q$}

\ared{As the volumetric source increases the stagnation surface $r=\sqrt{Q}$ migrates far from the origin. As such, taking $\sqrt{Q} \to \infty$ one expects the stagnation surface to be locally planar, valid, of course, for distances from the stagnation surface small compared with its radius $\sqrt{Q}$. In this vicinity, the velocity field~\eqref{vrvtvz} reduces, in the first approximation, to
\begin{equation}
    v_x = -2x, \quad v_y = \frac{R}{2}\mathrm{erfc}(-x), \quad v_z = 2 z \label{vxvyvz}
\end{equation}
where we have defined the local coordinates 
\begin{equation}
    x=r-\sqrt Q, \quad y =  \sqrt Q \theta, \quad  z=z  \label{xyz}
\end{equation}
such that $(x,y,z)\ll \sqrt{Q}$. \red{To derive $v_y$, we need to use Tricomi's expansion~\cite{tricomi1950asymptotic} for the incomplete gamma function.} The velocity field~\eqref{vxvyvz}, referred in the literature as the Burgers' vortex sheet \cite{burgers1948mathematical}, is also an exact solution of the Navier-Stokes equations. The vorticity component (equation~\eqref{wz}) decays like $1/\sqrt Q$ (so does $v_\theta$) and thus when rescaled, the vorticity component in the new coordinate system (using the same symbol) becomes
\begin{equation}
    \omega_z = \frac{R}{\sqrt{\pi}}e^{-x^2}. \label{wz2}
\end{equation}
The only parameter characterizing the flow here is thus the vortex sheet strength $R$.
}

\subsection{Generalized Sullivan-type vortices in radial stagnation flows}
\ared{
Sullivan \cite{sullivan1959two} extended the classical Burgers' solution by including radial variations in the axial velocity. The existence of a generalized Burgers' vortex in the presence of volumetric source, described above, suggests that a similar solution would hold for a vortex of the Sullivan type. As such, one may consider a velocity field of the form 
\begin{equation}
    v_r = -r + \frac{Q}{r}+ \frac{f(r)}{r}, \quad v_\theta=\frac{R}{r}g(r), \quad v_z = 2z - \frac{z}{r}\frac{df}{dr}.
\end{equation}
Evidently when $Q=0$ the velocity field must reduce to that given by Sullivan \cite{sullivan1959two}. The resulting solution and characteristics are described further in Appendix B so that our focus may return to the Burger-type vortex and its stability. Interesting enough, when $Q=6$, the Sullivan-type vortex and the Burgers-type vortex become identical.}


\section{Two-dimensional stability of the axisymmetric vortex}
In the case of Burgers' vortex ($Q=0$), the vortex system is found to be stable for infinitesimal disturbances lying in a plane \red{($r\theta$-plane)} perpendicular to the symmetry axis~\cite{robinson1984stability,prochazka1995two}; in particular, the vortex system is found to be more stable as the vortex Reynolds number $R$ increases, indicating that the swirling motion in this case essentially has stabilizing characteristics. The stability aspects when disturbances lie in an axial plane \red{($rz$-plane)} are not so trivial~\cite{crowdy1998note}, although, however, the vortex system is found to be asymptotically stable for small three-dimensional disturbances~\cite{schmid2004three}. The influence of the presence of $Q$ in~\eqref{vrvtvz} on the stability characteristics is investigated here for the two-dimensional disturbances lying in a plane perpendicular to the symmetry axis. The solution method thus follows Robinson and Saffman~\cite{robinson1984stability} and Prochazka and Pullin~\cite{prochazka1995two} closely.

Consider infinitesimal perturbations superposed upon the base flow so that the velocity takes the form
\begin{equation}
    \begin{bmatrix}
           -r+Q/r \\
           Rg(r)/r \\
           2z
         \end{bmatrix} +
         \begin{bmatrix}
           u'(r,\theta,t) \\
           v'(r,\theta,t) \\
           0
         \end{bmatrix}.
\end{equation}
The perturbations are easily described in the vorticity-stream function formulation where we define a perturbation stream function $\psi'$ such that
\begin{equation}
    u'= \frac{1}{r}\pfr{\psi'}{\theta}, \quad v'=-\pfr{\psi'}{r}.
\end{equation}
The corresponding non-zero vorticity component lying along the $z$ direction is given by $\omega'=-\nabla^2\psi'$. Introducing the normal mode decomposition
\begin{equation}
    \psi' = \psi(r) e^{-\mu t +in\theta} \qquad \omega' = \omega(r) e^{-\mu t +in\theta} \label{normalmodes}
\end{equation}
into the governing equations, the required linearized equations may be written as
\begin{align}
    M^n\psi&=-\omega , \label{Meq}\\
    L^n\omega +\mu\omega &= inR(g\omega+ f\psi )/r^2 \label{Leq}
\end{align}
where $f= (Q-r^2)r^Q e^{-r^2/2}/[2^{Q/2}\Gamma(1+Q/2)]$. In writing the above equations, we have defined the differential operators
\begin{align}
    M^n&\equiv \frac{1}{r}\frac{d}{dr}\left(r\frac{d}{dr}\right)-\frac{n^2}{r^2}, \label{Mpsi}\\
    L^n&\equiv \frac{1}{r}\frac{d}{dr}\left(r\frac{d}{dr}\right) + \left(r-\frac{Q}{r}\right)\frac{d}{dr}+ 2-\frac{n^2}{r^2}. \label{Lomeg}
\end{align}
Since our primary interest lies in the real part of $\mu$, without loss of generality, we can take $n\geq 0$ since the transformation $n\rightarrow -n$ is found to be equivalent to $\Im(\mu)\rightarrow -\Im(\mu)$. This point will be justified below, following equation \eqref{eig}. The required boundary conditions for the linearized problem are dictated by the conditions that the disturbances must decay as $r\rightarrow \infty$ and are regular at the origin, in particular the vorticity perturbations must decay exponentially as $r\rightarrow \infty$. This will become clearer after we solve the $R=0$ case explicitly.

 \subsection{The case $R=0$}
 When $R=0$, it is enough to consider the equation $L^n\omega + \mu\omega=0$ because $\omega$ decouples from $\psi$. The origin is a regular singular point of the operator $L^n$ where the local behaviour of the vorticity equation can be written as
\begin{equation}
    \omega \sim r^{\alpha+Q/2} \label{w0}
\end{equation}
where $\alpha=\sqrt{n^2+Q^2/4}$. The full solution of $\omega$ which is regular at the origin can be written, at once, in terms of the confluent hypergeometric function $M$ as follows 
\begin{equation}
    \omega = r^{\alpha+Q/2} M\left(\frac{\mu+\alpha+Q/2+2}{2},\alpha+1,-\frac{r^2}{2}\right). \label{wM}
\end{equation}
As expected, this solution reduces to that found by~\cite{robinson1984stability} for $Q=0$. The asymptotic expansion of this solution as $r\rightarrow \infty$ has an algebraically decaying part that decays like (cf. eqn. 2.5 in~\cite{robinson1984stability})
\begin{equation}
    \omega \sim r^{-\mu-2}/\Gamma(\alpha/2-Q/4-\mu/2) \label{winfal}
\end{equation}
and an exponentially decaying part given by 
\begin{equation}
    \omega \sim r^{\mu+Q}e^{-r^2/2}. \label{winf}
\end{equation}
The algebraically decaying part cannot be interpreted as a small perturbation to the \red{exponentially-decaying} base flow even when $R=0$ \red{(Localized disturbances are expected to decay rapidly outside the mixing layer)}. Requiring that it vanishes selects a discrete set of values for $\mu$, associated with the Gamma function in \eqref{winfal}, namely
\begin{equation}
    \mu_k = 2k+\alpha-Q/2, \qquad k=0,1,2,3,\dots \label{muk}
\end{equation}
Since $\alpha\ge Q/2$, it is clear that $\mu_k\ge0$. As such, for the case $R=0$, the flow is stable -- marginally when both $n = k = 0$ and asymptotically otherwise. In the former case, the perturbations \eqref{normalmodes} are stationary, with \eqref{wM} [or \eqref{omegk} below] reducing to \eqref{wz}. In this special case the perturbation takes the form of the new vortex solution described in the previous section but with small amplitude.

We note here that because of the way $n$ appears on the right hand side of \eqref{Leq} and in the operator \eqref{Lomeg}, that the stable solution described in the previous paragraph is also the solution for non-zero values of $R$ when $n=0$. For this reason we need only to discuss values of $n>0$ in what follows. 


Substituting~\eqref{muk} into~\eqref{wM}, the eigenfunctions corresponding to each $\mu_k$ can be expressed using the generalized Laguerre polynomials $L_k^{(\alpha)}$ as 
\begin{align}
    \omega_k  = Ar^{\alpha+Q/2}e^{-r^2/2}L_k^{(\alpha)}(r^2/2). \label{omegk}
\end{align}
where $A=\{k!/[2^\alpha\Gamma(\alpha+k+1)\}^{1/2}$. These eigenfunctions form a complete orthonormal set. With the appropriate inner product definition,
\begin{equation}
    \langle u,v\rangle = \int_0^\infty u \bar v r^{1-Q}e^{r^2/2}dr
\end{equation}
the orthogonality condition can be written as $\langle\omega_j,\omega_k\rangle=\delta_{jk}$. The streamfunction can be calculated by inverting the operator $M^n$ on $-\omega$.  For $n\geq 1$, this leads to
\begin{align}
    \psi_k = &\frac{Ar^n}{2n}\left[\int_0^\infty s^{\alpha+Q/2-n+1}e^{-s^2/2}L_k^{(\alpha)}(s^2/2)ds - \int_0^r s^{\alpha+Q/2-n+1}e^{-s^2/2}L_k^{(\alpha)}(s^2/2)ds\right]\nonumber \\ + 
     &\frac{Ar^{-n}}{2n} \int_0^r s^{\alpha+Q/2+n+1}e^{-s^2/2}L_k^{(\alpha)}(s^2/2)ds.
\end{align}
 The integrals appearing in the foregoing equation can be expressed in terms of generalized hypergeometric function of the type ${}_2 F_2$.

\subsection{Solution for general values of $R$}

Let us define the operator
\begin{equation}
    \mathcal{L}^n \equiv L^n - inRr^{-2}[g - f (M^n)^{-1}]
\end{equation}
so that the required eigenvalue problem for $R\neq 0$ is reduced to the single equation $\mathcal{L}^n\omega + \mu\omega=0$. A local analysis of this equation shows that $\omega$ retains the behaviour ~\eqref{w0} and~\eqref{winf}, respectively as $r\rightarrow 0$ and $r\rightarrow\infty$ when $R$ is non-zero. This enables us to expand $\omega$ in terms of the eigenfunctions~\eqref{omegk} identified earlier. Thus we shall take
\begin{equation}
    \omega = \sum_{k=0}^{N-1} a_k \omega_k \label{expan}
\end{equation}
where the series is truncated after the first $N$ terms for the purpose of numerical integrations.

\begin{figure}
\centering
\includegraphics[width=0.6\textwidth]{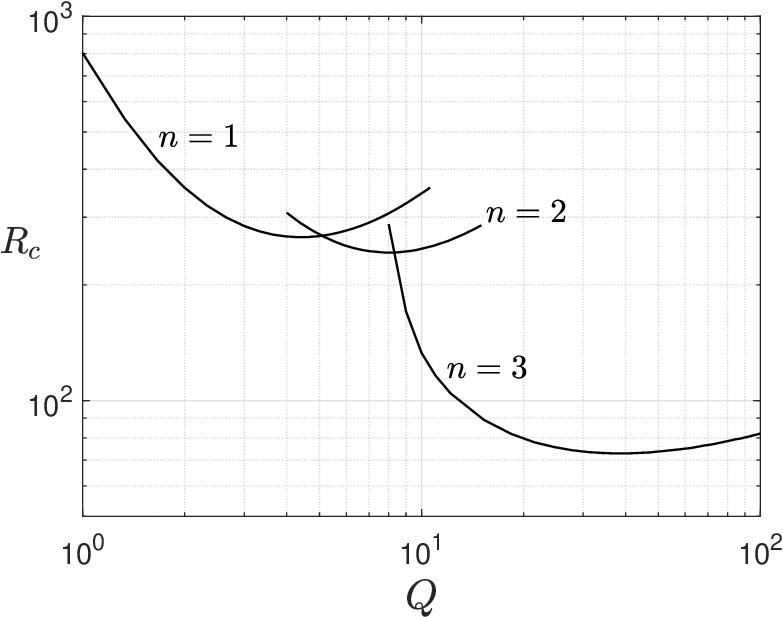}\\
\caption{The curves of critical Reynolds number for $n=1$, $n=2$ and $n=3$.}
\label{fig:rec}
\vspace{-0.15in}
\end{figure}

Substituting~\eqref{expan} into $\mathcal{L}^n\omega + \mu\omega=0$ and utilizing the orthogonality condition leads to an $N\times N$ matrix eigenvalue equation
\begin{equation}
    \vect A \vect x =  \mu \vect x \label{eig}
\end{equation}
where $\vect x=(a_0,a_1,a_2,\cdots)$ and $A_{jk} = -\langle \omega_j,\mathcal{L}^n\omega_k\rangle=\delta_{jk}\mu_k + inR\langle\omega_j,(g\omega_k + f\psi_k)/r^2\rangle$. Because $\mu_k$ is real, the imaginary part of $A_{jk}$, proportional to $n$, originates from the second term in this expression. Thus while the transformation $n \to - n$, leads to a change in sign of $\Im(A_{jk})$ and $\Im(\mu)$, the real part of $\mu$ is unchanged. As such, a full description of the stability is provided by considering non-negative values of $n$ only.

The eigenvalue equation~\eqref{eig} is solved numerically with $N=150$. This value for $N$ is found to be sufficient for Reynolds numbers less than $1000$. For each non-zero value of $Q$, the real part of $\mu$ is found to become negative, above a critical Reynolds number denoted here by $R_c$, for $n\geq 1$. The critical Reynolds number $R_c$ as a function of $Q$ is shown in Figure~\ref{fig:rec} for the first three azimuthal modes. For each azimuthal mode $n$, $R_c$ has its minimum value for a particular value of $Q$. For small values of $Q$, corresponding to the stagnation surface $r = \sqrt{Q}$ lying close to the axis, the least stable azimuthal mode corresponds to $n=1$, associated with a perturbation wavelength $\lambda= 2\pi\sqrt Q/n$. For very small values of $Q$, the system is stable since $R_c$ \red{is expected to become unbounded} as $Q$ \red{approaches either zero or some small finite value}. This agrees with the predictions given in \cite{robinson1984stability} and \cite{prochazka1995two}. It is clear from the figure that the envelope formed by the curves from the bottom side corresponding to different azimuthal modes determines the ultimate stability curve with the region lying below this curve corresponding to a stable regime and the region lying above corresponding to an unstable regime. 

\red{As $\sqrt Q \rightarrow \infty$, the velocity field approaches the Burgers' vortex sheet~\eqref{vxvyvz}, the stability analysis of which, has been addressed by Berenov and Kida~\cite{beronov1996linear}. It can be shown that the linearized equations~\eqref{Meq} and~\eqref{Leq}, as $\sqrt Q$ becomes large, reduce to equation (9) in~\cite{beronov1996linear}.  The critical curve $R_c$ as a function of $\lambda^{-1}$ indicates that $R_c$ decreases monotonically as $\lambda^{-1}$ decreases and exhibits the asymptotic behavior, $R_c = 4\sqrt 2 + 24 \sqrt \pi \lambda^{-1} + O(\lambda^{-2})$ as $\lambda^{-1}\rightarrow 0$.\footnote[1]{\red{The following correspondences should be made between the parameters used here and in~\cite{beronov1996linear} (we shall add to these parameters a subscript BK) before interpreting their results: strain rates are related by $A_{\mathrm{BK}}=2k$, vortex-sheet strengths are related by $\Gamma_{\rm{BK}}=\Gamma \sqrt k/\sqrt{2\nu}$, Reynolds numbers by $R_{\mathrm{BK}}=R/(4\sqrt 2)$ and the perturbation wavenumbers by $\alpha_{\mathrm{BK}}=n/(2\sqrt Q)=\sqrt 2\pi/\lambda$.}} This result holds so long as the planar approximation remains valid described by the condition that $\lambda \ll \sqrt Q$ or in other words $n$ must be such that $1\ll n\ll \sqrt Q$. In the first approximation, the shear layer becomes unstable when $R>4\sqrt 2$ so that the ultimate stability curve in Figure~\ref{fig:rec} asymptotes to this number for large $Q$. The first correction arising from the dependence on $n$ indicates that the least unstable mode, for a given $Q$, occurs for smaller values of $n$ satisfying the condition mentioned above. For sufficiently large $Q$, one should always be able to find values of $n$ such that $R_c$ is very close to $4\sqrt 2$. This means that, in practice, a wide range of azimuthal modes are admissible.}

\section{Conclusions}
A two-parameter family of exact solution of the Navier-Stokes equations is described that represents steady, axisymmetric vortices superposed on radial stagnation flows. The vortices are, in general, stable for small values of the vortex Reynolds number $R$ and the line-source Reynolds number $Q$. In tubular-flame experiments or flagpole burning in a fire  whirl, both of these Reynolds number are expected to be large, in which case, the vortex system may destabilize and can undergo transition to another state or, turbulence. \red{At least, in the planar case, we know that after the initial Kelvin-Helmholtz instability of the strained shear layer, secondary instabilities lead to counter-rotating vortices~\cite{neu1984dynamics,lin1984mixing}, for which a steady-state solution exists, as discovered by Kerr and Dold~\cite{kerr1994periodic}. A similar transition and the existence of steady counter-rotating vortices lying along the stagnation surface can be expected for radial stagnation flows also. Moreover, by conformally mapping the Burgers' vortex-layer solution, Bazant and Moffatt~\cite{bazant2005exact} found additional solutions of various types. Such solutions also may exist in cylindrical stagnation flows. Further, one may be able to identify unsteady solutions subjected to time-varying strain rates and source strengths, similar to that identified for Burgers' and Sullivan vortices (a good summary of these unsteady solutions is given in~\cite{drazin2006navier}).}

\red{The other interesting aspect of the problem concerns the Sullivan-type vortex, described in Appendix B. Although whether one observes either the Burgers-type vortex or the Sullivan-type vortex as a steady-state vortex, depends on how the steady state is established, the existence of a bifurcation between the two types of vortices at $Q=6$ is a plausible expectation and merits consideration in the future. All these observations evidently point to a fertile field for future investigations.} 


\section{Acknowledgments}

The authors are grateful to Forman A. Williams and Antonio L. S\'anchez for their insightful discussions regarding the problem addressed here. \red{The authors are also thankful to two anonymous reviewers for their helpful comments.}

\section*{Appendix A}
 Here we generalize the vortex solution described in section  \ref{sec:vortexsoln} by replacing the line source at the axis with a porous cylinder of finite radius. Following~\cite{wang1974axisymmetric,cunning1998radial}, we can introduce the substitution
\begin{equation}
    v_r = - \frac{1}{r}f(\eta), \quad v_\theta = \frac{R}{r}g(\eta), \quad v_z = 2z\frac{df}{d\eta}
\end{equation}
where $\eta=r^2$, into the governing equations. The resultant equations then become
\begin{align}
   \frac{d}{d\eta}\left(2\eta\frac{d^2f}{d\eta^2}\right)  + f\frac{d^2f}{d\eta^2} - \left(\frac{df}{d\eta}\right)^2 +  1   =0, \label{feq}\\
    2\eta \frac{d^2g}{d\eta^2} +  f \frac{dg}{d\eta} =0. \label{geq}
\end{align}
The required boundary conditions for our case are given by
\begin{align}
    f + Q= \frac{df}{d\eta}=g=0\quad \text{at} \quad \eta=\eta_w,\\
    \frac{df}{d\eta}=g= 1 \quad \text{as} \quad \eta\rightarrow \infty.
\end{align}
where $Q=R_1U_1/\nu$ and $\eta_w=kR_1^2/\nu$. 
\begin{figure}
\centering
\includegraphics[width=0.6\textwidth]{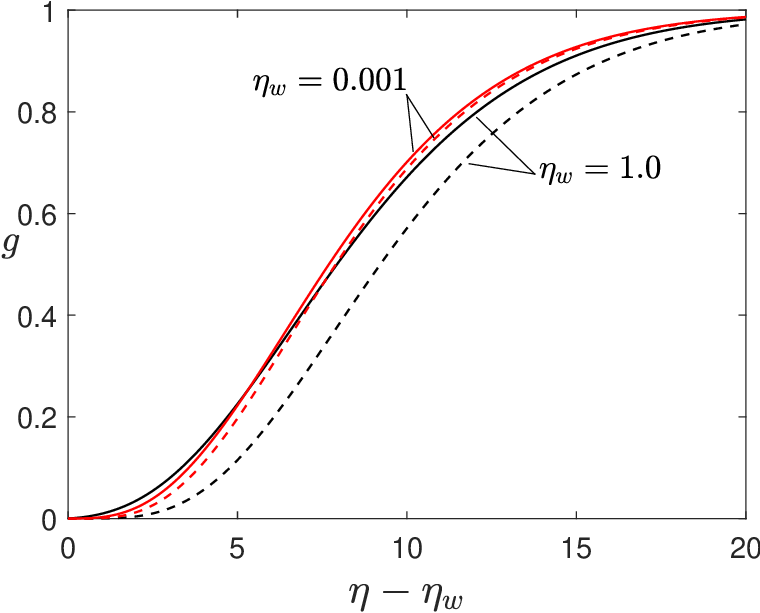}\\
\caption{The curves of $g(\eta)$ for two different values of $\eta_w$ with solid curves corresponding to the solution of~\eqref{geq} with $Q=4$ and dashed curves representing the expression~\eqref{gsol} calculated with corresponding values of $Q_v$.}
\label{fig:g}
\vspace{-0.15in}
\end{figure}

The vortex solution derived in the main text emerges for the range $\eta\gg \eta_w$. In this range, equation~\eqref{feq} admits the solution $f= \eta-Q_v$ (equivalent to~\eqref{vrvtvz} but $Q$ replaced with $Q_v$), where $Q_v$ is termed as the virtual flow rate. The virtual flow rate is the flow rate that would be needed if the solution for $f$ is everywhere linear (of course, being a potential solution, the linear form do not satisfy the slip condition at the wall). A familiar example that is
analogous to this situation is the concept of displaced origin in the description of finite-size effects of the self-similar solutions of jets and plumes~\cite{da1937velocity,revuelta2002virtual}.  The value of $Q_v$ can be extracted from the numerical solution of~\eqref{feq} by taking the limit $Q_v=\lim_{\eta\rightarrow\infty}(\eta-f)$.

Since vorticity tends to concentrate near the stagnation surface, the range of validity of vortex solution can written as $\eta_s(\eta_w,Q)\gg \eta_w$, where $f(\eta_s)=0$. Numerical integrations of $f(\eta)$ suggested that $\eta_s$ depends weekly on $\eta_w$ and that $\eta_s\sim Q$ (the stagnation location defined in the introduction) as far as orders of magnitude are concerned. The required condition then becomes $Q\gg \eta_w$, or, in terms of dimensional variables $U_1\gg kR_1$. The function $g(\eta)$ calculated from~\eqref{geq} is shown in Figure~\ref{fig:g} for two different values of $\eta_w$ and for a given value of $Q$. The result is compared with the expression~\eqref{gsol}
 with $Q_v$ replacing $Q$. As can be seen, in the case $Q\gg \eta_w$, the agreement between the two forms of $g$ is good.

\section*{Appendix B}
\begin{figure}
\centering
\includegraphics[width=1.0\textwidth]{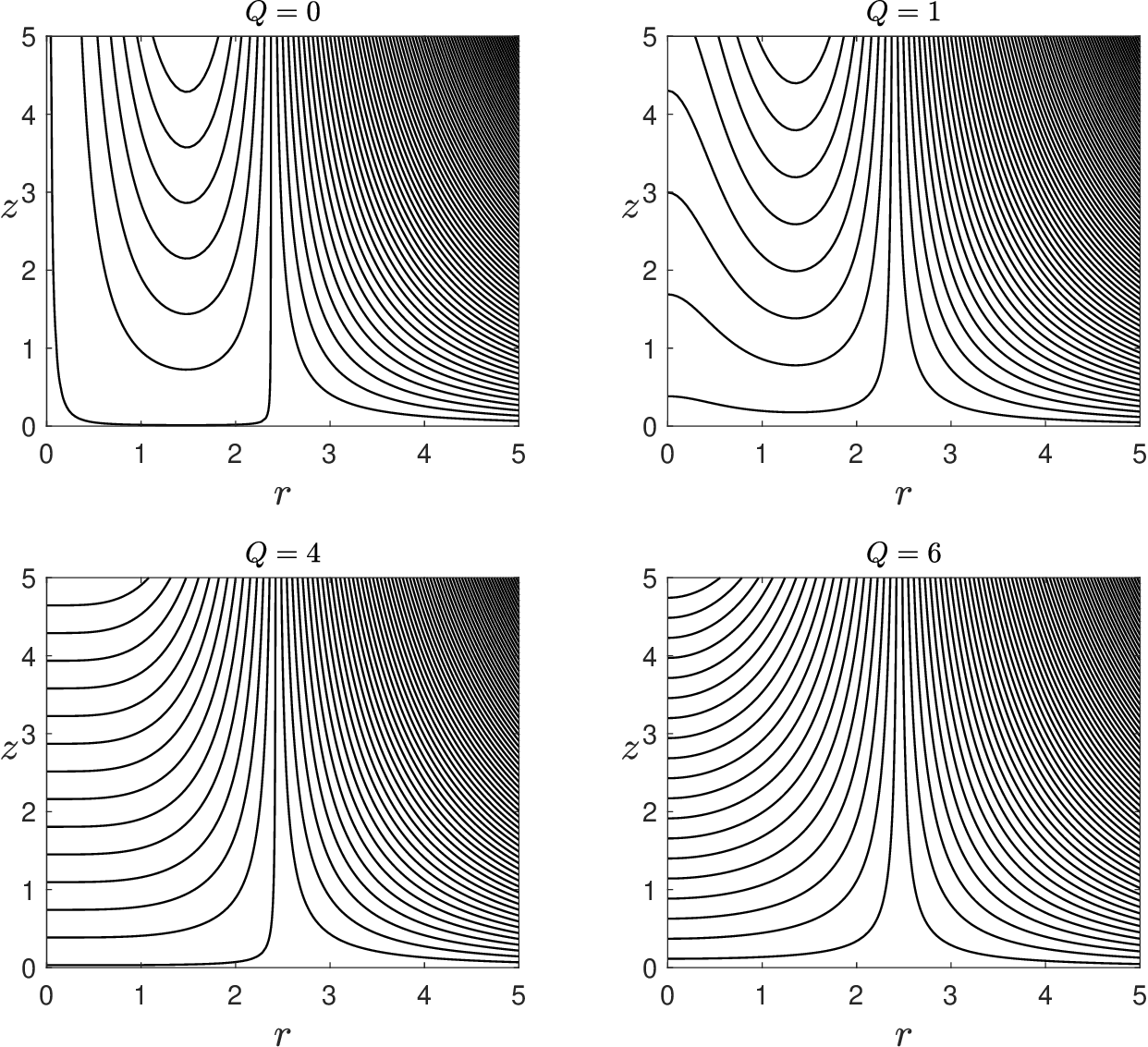}\\
\caption{Streamlines in the axial plane of the Sullivan-type flow field for $Q=(0,1,4,6)$.}
\label{fig:sullivan}
\vspace{-0.15in}
\end{figure}
\red{In the main text, Burgers' solution was extended to include nonzero values of $Q$. Here the Sullivan-type vortex is similarly treated. Following~\cite{sullivan1959two}, consider
\begin{equation}
    v_r = -r + \frac{Q}{r}+ \frac{f(r)}{r}, \quad v_\theta=\frac{R}{r}g(r), \quad v_z = 2z - \frac{z}{r}\frac{df}{dr}. \label{vrvtvzsull}
\end{equation}
 This leads us to the following solution
\begin{align}
    f(r) &= (6-Q)(1-e^{-r^2/2}),\\
    g(r) &= A \int_0^r s^7 e^{-s^2/2}\exp[(Q/2-3)\mathrm{Ei}(-s^2/2)]ds \label{gsull}
\end{align}
where
\begin{equation}
    1/A = \int_0^\infty s^7 e^{-s^2/2}\exp[(Q/2-3)\mathrm{Ei}(-s^2/2)]ds
\end{equation}
and $\mathrm{Ei}$ is the exponential integral. It is evident that when $Q=6$ the Sullivan-type vortex and the Burgers-type vortex become identical.}

\red{In the Burgers-type vortex, the cylindrical stagnation surface is located at $r=\sqrt Q$, whereas in the Sullivan-type vortex, the radial location of the stagnation surface is given by
\begin{equation}
    r =  \sqrt{6+2W_0\left(\frac{Q-6}{2e^3}\right)} \label{rsull}
\end{equation}
where $W_0$ represents the principal branch of the Lambert W function. Although in the main text we are interested in values of $Q\geq 0$, it is worthwhile to point out that the Sullivan-type vortex solution continues to exist for values $-2e^2 + 6\leq Q<0$. In that range, there are two stagnation surfaces, one (with flow approaching towards the surface) corresponding to the aforementioned formula and the second (with flow diverging away from it) corresponding to \eqref{rsull} with $W_0$ replaced by the lower branch $W_{-1}$. The two stagnation surfaces approach each other reaching their limiting value $r=2$, as $Q\rightarrow -2e^2 + 6$. Whenever $Q<0$, the vorticity is singular at the axis.}

\red{In the Burgers-type vortex, $v_z/z>0$ for all values of $r$. In the Sullivan-type vortex, we have, for $ Q<4$
\begin{align}
    v_z/z<0 \quad \text{for} \quad r^2<2\ln\left(\frac{6-Q}{2}\right)\\
    v_z/z>0 \quad \text{for} \quad r^2>2\ln\left(\frac{6-Q}{2}\right)
\end{align}
and for $Q>4$, $v_z/z>0$ for all values of $r$. Since $v_z=0$ at $r=0$ for $Q=4$, the streamlines emerge from the axis in purely radial direction. These observations are more apparent in the streamline plots shown in Figure~\ref{fig:sullivan} for $z>0$. As $Q\rightarrow \infty$, the W function increases very moderately (i.e., $W_0(x)\rightarrow \ln(x)$) such that the stagnation location~\eqref{rsull} stays order unity for large values of $Q$, with the result that equations~\eqref{vrvtvzsull}-\eqref{gsull} remain the appropriate description for large $Q$. }

\bibliographystyle{unsrt}
\bibliography{references}

\end{document}